\documentclass[a4paper,twoside]{article}
\usepackage{richpom,graphicx}
\begin{document}
\title{EINSTEIN-MAXWELL FIELDS GENERATED FROM THE $\gamma$-METRIC AND THEIR LIMITS}
\author{%
L. Richterek\footnote{E-mail: {\tt richter@aix.upol.cz}}\\
    {\sl Department of Theoretical Physics, Palack\'y University,}\\
    {\sl 17. listopadu 50, Olomouc, 772 00}\\[\medskipamount]
J. Novotn\'y\footnote{E-mail: {\tt novotny@physics.muni.cz}}\\   
    {\sl Department of General Physics, Masaryk University}\\[\medskipamount]  
J. Horsk\'y\footnote{E-mail: {\tt horsky@physics.muni.cz}}\\   
    {\sl Institute of Theoretical Physics and Astrophysics, Masaryk University,}\\ 
    {\sl Kotl\'a\v{r}sk\'a 2, Brno, 611 37, Czech Republic}
}\maketitle

\begin{abstract}
Two solutions of the coupled Einstein-Maxwell field equations are found by
means of the Horsk\'y-Mitskievitch generating conjecture. The vacuum limit
of those obtained classes of spacetimes is the seed $\gamma$-metric and
each of the generated solutions is connected with one Killing vector of the seed
spacetime. Some of the limiting cases of our solutions are identified with
already known metrics,  the relations among various limits are
illustrated through a limiting diagram. We also verify our calculation
through the Ernst potentials. The existence of circular geodesics is briefly
discussed in the Appendix.
\end{abstract}

\section{Introduction}
Several constructing techniques has been proposed that suggest how to
generate Einstein-Maxwell (E-M) fields from pure gravitational ones (see
e.g. \cite{KSMH}). Hereafter in this paper we use an alternative method
based on the Horsk\'y-Mitskievitch (H-M) conjecture \cite{horsky:1989} which
prescribes quite close connection between isometries of vacuum spacetimes
(seed metrics) and an electromagnetic four-potential of generated E-M
fields. Taking the $\gamma$-metric as a seed vacuum spacetime, we obtain two
classes of E-M fields, each of which corresponds to one Killing
vector of the seed metric. Main properties of those solutions are summarized too.

The paper is organized in the following way: we start resuming the basic
characteristic of the $\gamma$-metric, then we gradually come to new classes
of E-M fields that further generalize the results presented in
\cite{richterek:2000}. We demonstrate that some already known E-M fields
can be regarded as special cases of these solutions. The most
important limits and special cases of the generated E-M fields are listed
in section~\ref{sec:limcases}, the relations among them are illustrated
through a limiting diagram. At the same time we touch on problems connected
with the physical interpretation of obtained solution, with the existence
of curvature singularities and circular geodesics.

\section{The seed $\gamma$-metric}
\label{sec:gammet}

The $\gamma$-metric is a vacuum solution of Einstein equations discovered
by Darmois in 1927 and reinvestigated by various authors many times since
(see references cited in \cite{bonnor:1992} for more details). Known also as
Darmois-Voorhees-Zipoy metric, it represents an interesting class of static
axially symmetric spacetimes. Therefore it can be expressed in the
Weyl-Lewis-Papapetrou cylindrical coordinates in the form (see e.g.~\cite{bonnor:1992})
\begin{equation}
{\rm d}s^{2}=-e^{2\mu}{\rm
d}t^{2}+e^{-2\mu}\left[e^{2\nu}\left({\rm d}r^{2}+{\rm
d}z^{2}\right)+r^{2}{\rm d}\varphi^{2}\right],
\label{eq:dsweyl}
\end{equation}
where $\mu =\mu (r,z),\ \nu=\nu (r,z)$ are functions of $r$ and $z$ only.
Particularly, the $\gamma$-metric is the line-element (\ref{eq:dsweyl}) with
\cite{herrera:1999} 
\begin{eqnarray}
{\rm e}^{2\mu} = 
\left(\frac{R_{1}+R_{2}-2m}{R_{1}+R_{2}+2m}\right)^{\gamma}=f_1(r,z),\nonumber\\
{\rm e}^{2\nu} = \left[\frac{(R_{1}+R_{2}-2m)(R_{1}+R_{2}+2m)}
        {4R_{1}R_{2}}\right]^{\gamma^{2}}=f_2(r,z),\label{eq:fcegam}\\
R_{1} = \sqrt{r^{2}+\left(z-m\right)^{2}},\quad
R_{2} = \sqrt{r^{2}+\left(z+m\right)^{2}}.\nonumber
\end{eqnarray}
It is well known that the function $\mu (r,z)$ satisfies flat-space
Laplace's equation $\Delta\mu =0$ for any Weyl metric~\cite{bonnor:1992}
and that it is possible to construct an infinite number axially symmetric
solutions of Einstein's equations using various realistic Newtonian
potentials identified with $\mu (r,z)$ \cite{bicak:2000}. Particularly, the
$\gamma$-metric can be generated by the Newtonian potential of a linear
mass source (``rod'') with linear density $\gamma/2$ and length
$2m$~\cite{herrera:1999}
\[
\phi=\mu (r,z) = \frac{1}{2}\ln \left|g_{tt}\right| = \frac{\gamma}{2} 
\ln\left(\frac{R_{1}+R_{2}-2m}{R_{1}+R_{2}+2m}\right).
\]
This fact is the main supporting argument for the ``standard'' interpretation
of the $\gamma$-metric as an exterior field of a finite linear source
located along the $z$-axis symmetrically with respect to the origin
$z=0$. However, though the function $\mu (r,z)$ may give us a guide to the
physical meaning of the exact solution, this correspondence has to be used
with caution; the Darmois-Voorhees-Zipoy solution can be also explained as a
gravitational field of counter-rotating relativistic
discs~\cite{bicak:2000,bicak:1993} or a field of an oblate ($\gamma >1 $)
or prolate ($\gamma <1 $) spheroid~\cite{herrera:1999}. Geodesic motion in
Darmois-Voorhees-Zipoy spacetime and its difference from the case of
spherical symmetry has been analyzed by Herrera, Paiva and
Santos~\cite{herrera:1998}.

The $\gamma$-metric has an interesting singularity structure: it has a
directional singularity for $\gamma > 2$, but not for $\gamma <
2$~\cite{bonnor:1992}. For a distant observer at infinity such
gravitational field behaves as an isolated body with monopole and higher
mass moments which can be -- at lest principally -- measured by gyroscope
experiments~\cite{herrera:2000}.  A superposition of two or more
$\gamma$-solutions was studied by Letelier and
Oliveira~\cite{letelier:1998}. Moreover, if
\begin{equation}
\lim_{r\to 0}\, \nu(r,z) \neq 0,
\label{eq-consing}
\end{equation}
then any Weyl metric~(\ref{eq:dsweyl}) includes a conical singularity
representing stresses on the $z$-axis (see
e.g.~\cite{bonnor:1992,bicak:2000}). 

The case $\gamma=1$ corresponds to the exterior Schwarzschild metric
outside the horizon. It can be more easily demonstrated when one uses
so-called Erez-Rosen spherical coordinates $\varrho$,
$\vartheta$~\cite{herrera:1999,herrera:1998} introduced by the
transformation
\begin{equation}
r^{2} = \left(\varrho^{2}-2m\varrho\right)\sin^{2}\vartheta,\quad
z=\left(\varrho-m\right)\cos\vartheta.
\label{eq:trerros}
\end{equation}
Naturally, our results concerning the $\gamma$-metric must be consistent with
this thoroughly studied Schwarzschild limit.

Recently, Herrera et al.~\cite{herrera:1999} showed that the limit
$m\to\infty$ locally gives the Levi-Civita (L-C) spacetime~\cite{bonnor:1992,wang:1997}
\begin{equation}
{\rm d}s^{2}=-r^{4\sigma}{\rm d}t^{2} + r^{4\sigma (2\sigma -
1)}\left({\rm d}r^{2}+{\rm d}z^{2}\right) + C^{-2}r^{2-4\sigma}{\rm
d}\varphi^{2},
\label{eq:dslc}
\end{equation}
i.e. another Weyl metric with constant parameters $\sigma,\ C$ (for the
relation between $\gamma$ and $\sigma$ in this limit see
section~\ref{sec:limcases}). We have already shown that every Killing
vector of the L-C spacetime is connected with a particular class of E-M
fields and that those fields can be generated through the H-M
conjecture~\cite{richterek:2000}. Naturally, a question arises, whether the
generating conjecture can be applied to the $\gamma$-metric in the same
simplified way as to the L-C metric. The existence of the solutions
described in section~\ref{sec:gammetelmg} proves that
for two Killing vectors $\partial_t$, $\partial_\varphi$ the answer is
positive.  
 
The H-M generating conjecture, which we apply to the $\gamma$-metric, seems
to outline an efficient and fruitful way, how to obtain solutions of E-M
equations as a generalization of some already known vacuum seed metrics,
though no general proof has been given so far and that we do not know exact
limits of its applicability (see~\cite{stephani:2000} and references
therein). In this paper we do not intend to tackle general theoretical
problems, we would like to concentrate on the application of the conjecture
to our particular case instead.

In~\cite{richterek:2000} we have proposed a simplified {\em ad hoc}
empirical algorithmic scheme, by means of which six classes of E-M fields
were obtained; each of those classes corresponds to another Killing vectors
of the seed L-C solution~(\ref{eq:dslc}). The fact that the L-C solution
represents a limiting case of the $\gamma$-metric provides us with the
motivation to apply this algorithm also in the case of Darmois-Voorhees-Zipoy
vacuum solution.

In the following section the H-M conjecture is applied to the seed
$\gamma$-metric in the same way we have applied it to the seed L-C
metric~\cite{richterek:2000}. Therefore we do not concentrate on the
reexplanation of the generation procedure, we rather describe basic
characteristics of the obtained solutions and then we study their limiting
cases (section~\ref{sec:limcases}). For our further calculations it is more
convenient to rewrite the line-element (\ref{eq:dsweyl}) in the form
\begin{equation}
{\rm d}s^{2} = -\,f_{1}(r,z)\,{\rm d}t^{2} + f_{1}(r,z)^{-1}\,
\left[f_{2}(r,z)\left({\rm d}r^{2}+{\rm d}z^{2}\right) + 
r^{2}\,{\rm d}\varphi^2\right],
\label{eq:dsgam}
\end{equation}
where the functions $f_{1}(r,z),\,f_{2}(r,z)$ are defined by (\ref{eq:fcegam}).

\section{The $\gamma$-metric with an electromagnetic field}
\label{sec:gammetelmg}
\subsection{The $\gamma$-metric with an electric field}
\label{sec:gammetel}
Let us first employ the timelike Killing vector
$\partial_t$. Following~\cite{richterek:2000}, we modify the
line-element~(\ref{eq:dsgam}) into the form
\begin{equation}
{\rm d}s^{2} = -\,\frac{f_{1}(r,z)}{f(r,z)^2}\,{\rm d}t^{2} + 
\frac{f(r,z)^2}{f_{1}(r,z)}\,
\left[f_{2}(r,z)\left({\rm d}r^{2}+{\rm d}z^{2}\right) + 
r^{2}\,{\rm d}\varphi\right],
\label{eq:dsgamel}
\end{equation}
and set the four-potential
\begin{equation}
\bsfvec{A}=q\,\frac{f_{1}(r,z)}{f(r,z)}\,\bsfvec{dt}.
\label{eq:vpgamel}
\end{equation}
It can be verified through standard calculations that sourceless E-M
equations are fulfilled if
\[
f(r,z)=1-q^2f_{1}(r,z).
\]
Generated E-M field is of an electric type, because of the non-positive
electromagnetic invariant
\[
\eqalign{F_{\alpha\beta}F^{\alpha\beta}=\\
\qquad\ =\,
-\,\frac{32\,q^2m^2\gamma^2f_{1}(r,z)^2
\left\{r^2\left(R_1+R_2\right)^2 + 
\left[z\left(R_1+R_2\right)+m\left(R_1-R_2\right)\right]^2\right\}}
{f(r,z)^4f_{2}(r,z)R_1^2R_2^2
\left( R_1+R_2-2m\right)^2\left( R_1+R_2+2m\right)^2} \leq 0.}
\]
This can be also checked when we calculate the non-zero components of the
electric field strength $E^{1} = F^{01}$ and $E^{3} = F^{03}$. 
The analytic expressions for the Kretschmann scalar as well as for the Weyl
scalars are too lengthy to reproduce them here in full extend. The
spacetime~(\ref{eq:dsgamel}) generally belongs to the Petrov type $I$, with the
trivial exception $\gamma=0$ when it belongs to the Petrov type $0$ and the
electromagnetic field disappears.

The spacetime~(\ref{eq:dsgamel}) has two curvature singularities. The
attractive one is inherited from the seed metric~(\ref{eq:dsweyl}) and its
location coincides with the Newtonian source -- a massive rod situated
along the $z$-axis symmetrically with respect to the origin.  The
Kretschmann scalar also blows up to infinity at ``points'', the coordinates
of which satisfy the transcendent equation
\begin{equation}
f(r,z)=1-q^2f_{1}(r,z) = 1 -
q^2\left(\frac{R_{1}+R_{2}-2m}{R_{1}+R_{2}+2m}\right)^{\gamma} = 0.
\label{eq:singgam}
\end{equation}
The presence and the ``shape'' of this latter repulsive curvature
singularity can be also illustrated by means of the corresponding generating
Newtonian potential extracted from the $g_{tt}$ component of the metric
tensor
\begin{figure}[t]
\parbox{0.45\textwidth}{
\centerline{\rotatebox{-90}{\includegraphics[width=5cm]{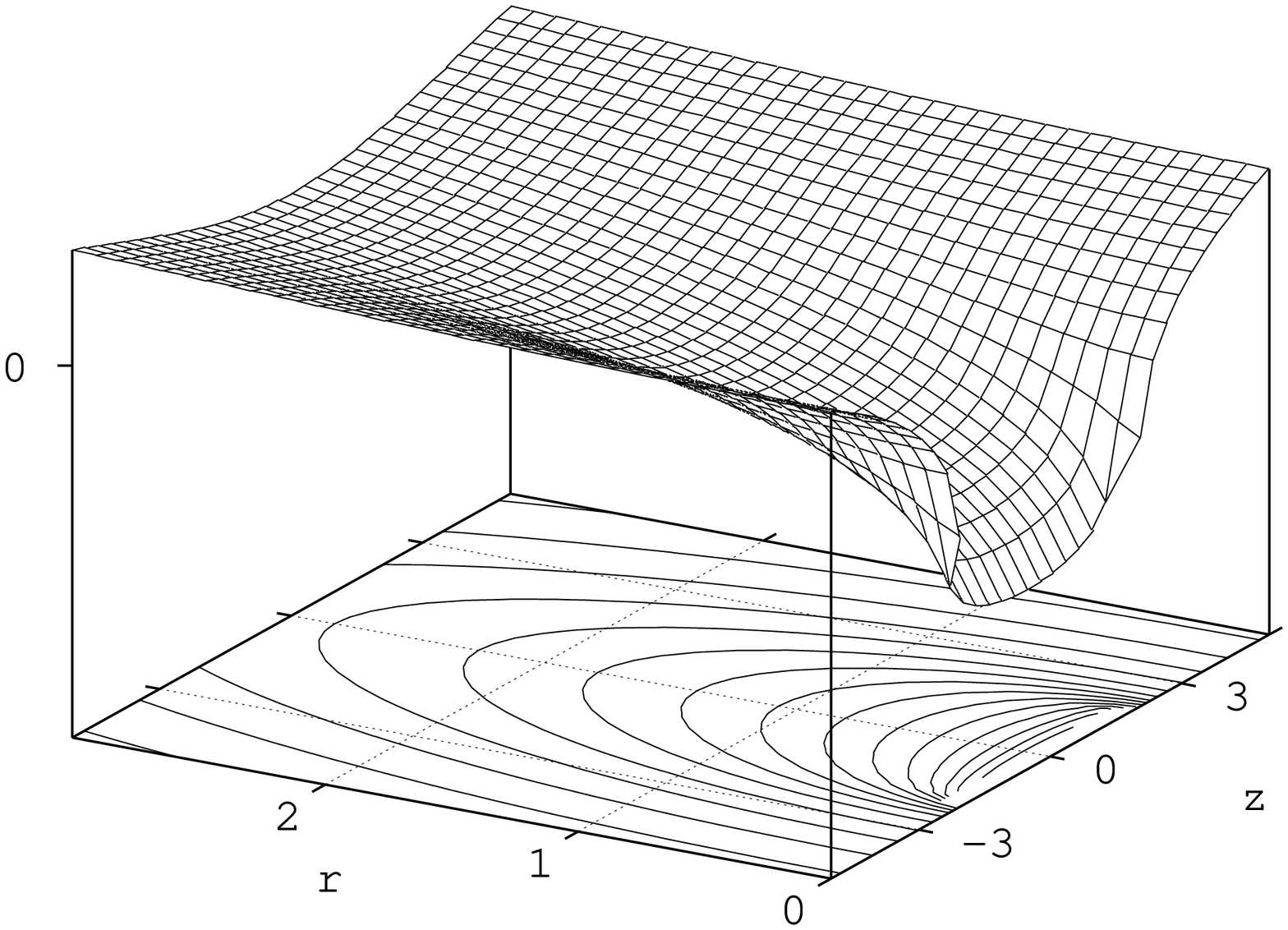}}}
\centerline{a)}
}
\hfill
\parbox{0.45\textwidth}{
\centerline{\rotatebox{-90}{\includegraphics[width=5cm]{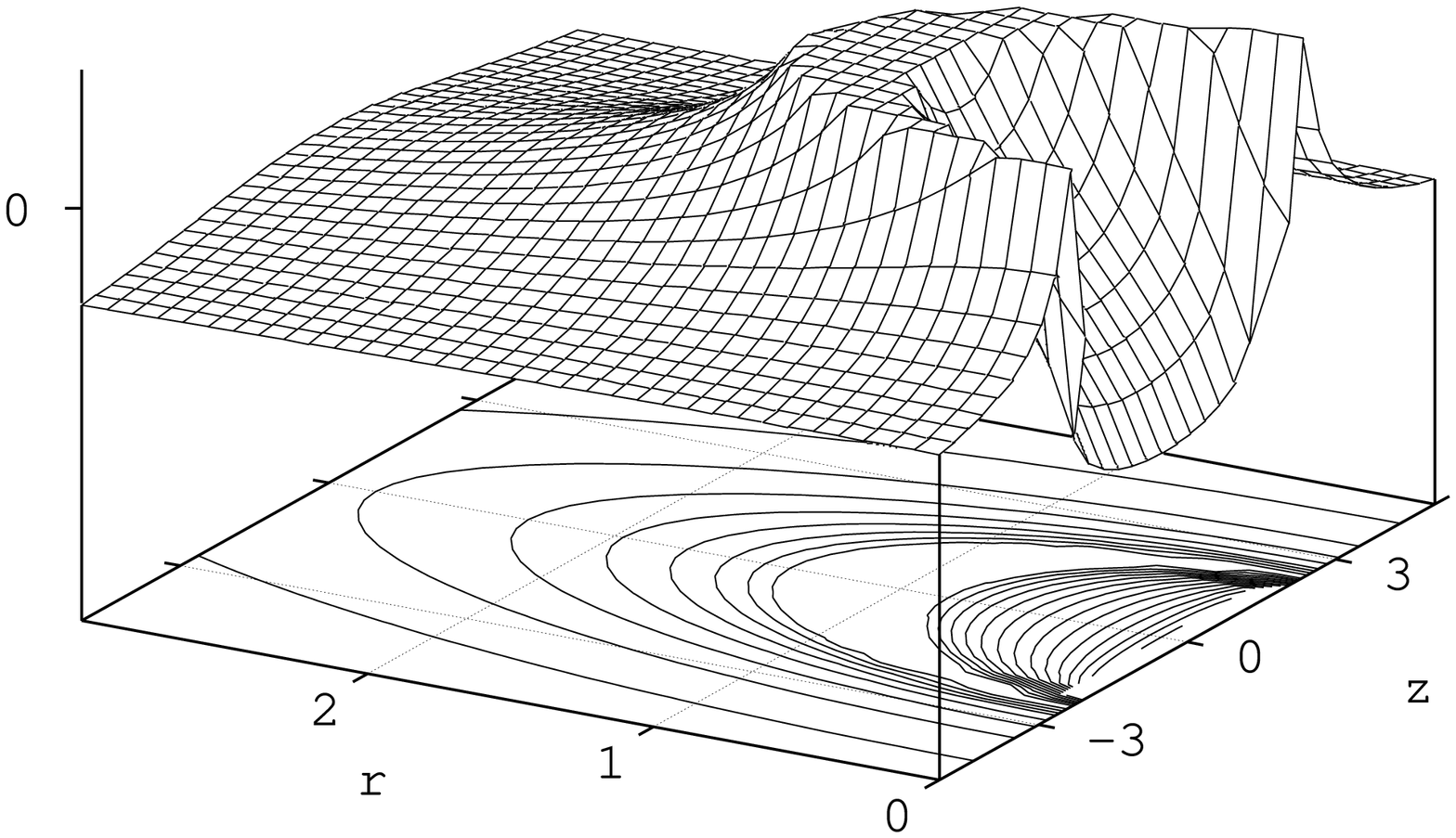}}}
\centerline{b)}
}

\vskip3ex
\centerline{
\parbox{0.45\textwidth}{
\centerline{\rotatebox{-90}{\includegraphics[width=5cm]{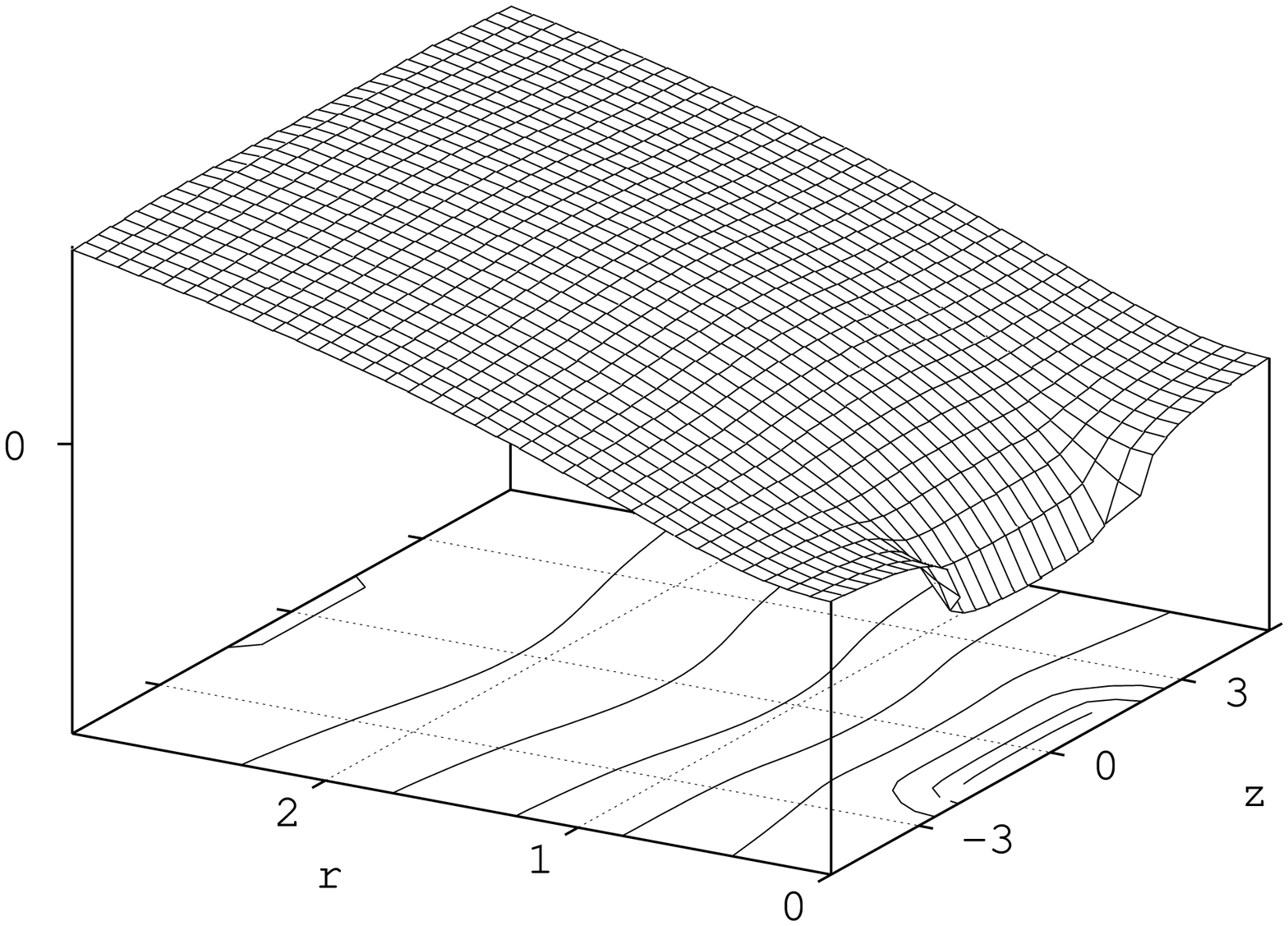}}}
\centerline{c)}
}}
\caption{Newtonian potential $\phi=\phi(r,z)$ for the ``charged''
$\gamma$-metric solutions with $m=2,\ \gamma=0.5$: 
a) the solution with an electric field, $q=1$;
b) the solution with an electric field, $q=2$; 
c) the solution with a magnetic field, $q=1$.}
\label{fig:newtgam}
\end{figure}
\begin{displaymath}
\phi = \frac{1}{2}\ln \left| g_{tt}\right| = \frac{1}{2}\ln
f_1\left(r,z\right) - \frac{1}{2}\ln\left[1-q^2f_1\left(r,z\right)\right]^2,
\end{displaymath}
the dependence of which on the coordinates $r,z$ together with the
equipotential curves are illustrated in figures~\ref{fig:newtgam}(a),
(b). While for some set of the parameters the condition~(\ref{eq:singgam})
cannot hold and the spacetime has only one rod-like singularity at $z$-axis
as in figure~\ref{fig:newtgam}(a), in other cases the spacetime is endowed
with a singularity surrounding the linear Newtonian source and acting as a
repulsing potential barrier in figure~\ref{fig:newtgam}(b).  At least part
of the electric field could be intuitively connected with the charge
distribution along the finite rod, but the physical character of the second
singularity present in this solution is not clear. It might seem that the
existence of such repulsive singularity excludes any ``reasonable'' physical
interpretation of our solution. On the other hand, as it is shown in
section~\ref{sec:limcases}, the metric (\ref{eq:dsgamel}) includes the
Reissner-Nordstr\"om solution as its limit. Therefore, the
Weyl-Lewis-Papapetrou coordinates are probably not the best suitable
coordinates for the study of this ``strange'' singularity. The existence of
circular geodesics is studied in the Appendix.

\subsection{The $\gamma$-metric with a magnetic field}
\label{sec:gammetmg}
Next, turn our attention to the Killing vector $\partial_\varphi$.  In
accordance with~\cite{richterek:2000} we assume the line-element
\begin{equation}
\eqalign{{\rm d}s^{2} =&-\,f(r,z)^2f_{1}(r,z)\,{\rm d}t^{2} +\\
&+\frac{1}{f_{1}(r,z)}
\left[f(r,z)^2f_{2}(r,z)\left({\rm d}r^{2}+{\rm d}z^{2}\right) + 
\frac{r^{2}}{f(r,z)^2}\,{\rm d}\varphi^2\right]}
\label{eq:dsgammg}
\end{equation}
and four-potential
\begin{equation}
\bsfvec{A}=q\,\frac{r^2}{f(r,z)f_{1}(r,z)}\,\bsfvec{d}\bvphi.
\label{eq:vpgammg}
\end{equation}
One can again verify the validity of sourceless E-M equations and the
condition of traceless Einstein tensor. Gradually, we get the result
\[
f(r,z)=1+q^2r^2/f_{1}(r,z). 
\]
The obtained electromagnetic field is of magnetic type because 
\begin{eqnarray*}
F_{\alpha\beta}F^{\alpha\beta)} =
\frac{32\,q^2}{f(r,z)^4f_{2}(r,z)R_1^2R_2^2
\left( R_1+R_2-2m\right)^2\left( R_1+R_2+2m\right)^2}\times\\
\times \left\{\left[R_1R_2\left(r^2+z^2-m^2\right) 
- m\gamma r^2\left(R_1+R_2\right)\right]^2\right. +\\
+\left. r^2m^2\gamma^2\left[R_2\left(z-m\right)+
R_1\left(z+m\right)\right]^2\right\} \geq 0;
\end{eqnarray*}
both longitudinal and radial components of the magnetic field strength
$B^{3} = F^{12}$, $B^{1} = F^{23}$ are nonzero.  The
Kretschmann and Weyl scalars are again too lengthy to reproduce them
here. Generated E-M field generally belongs to the Petrov type $I$ besides
the trivial exception $\gamma=0$ in which case we get the Petrov class $D$.

The solution~(\ref{eq:dsgammg}) can be evidently written in the Weyl
form~(\ref{eq:dsweyl}) and thus it can be considered as a spacetime generated
by the Newtonian gravitational potential
\begin{displaymath}
\phi = \frac{1}{2}\ln \left| g_{tt}\right| = \frac{1}{2}\ln
f_1\left(r,z\right) + \ln\left[1+\frac{q^2r^2}{f_1\left(r,z\right)}\right];
\end{displaymath}
the dependence of $\phi$ on Weyl coordinates $r,\ z$ is drawn in
figure~\ref{fig:newtgam}(c), from which the presence of rod-like curvature
singularity along $z$-axis can be deduced. This curvature singularity is,
of course, inherited from the seed $\gamma$-metric. 

\subsection{Ernst potentials and dual solutions}
\label{sec:ernstpot}
The coupled E-M filed equations for any axially symmetric stationary metric
\begin{equation}
{\rm d}s^{2}=-F(r,z)\,\left({\rm d}t- \omega_j{\rm d}x^j\right)^2
+F(r,z)^{-1}\left[e^{2\nu}\left({\rm d}r^{2}+{\rm
d}z^{2}\right)+r^{2}{\rm d}\varphi^{2}\right],
\label{eq:axisym}
\end{equation}
are often formulated in terms of two complex functions, so called Ernst
potentials ${\cal E}$ a $\Phi$~\cite{ernst:1968a,ernst:1968b} (an index
$j=1,2,3$ runs over space-like coordinates). As both
solutions~(\ref{eq:dsgamel}) and~(\ref{eq:dsgammg}) are static
(i.e. $\omega_j=0\ \forall j$), in their
case the Ernst potentials take a simplified form
\begin{eqnarray*}
\Phi &=& A_{t} + {\rm i}\,A_{\varphi}'\\
{\cal E} &=& F(r,z) - \left|\Phi\right|^2
\end{eqnarray*}
where $A_{\varphi}'$ depends on the vector potential components $A_{t}$
$A_{\varphi}$ via equation
\begin{equation}
\frac{1}{r}\,F(r,z){\nabla}A_{\varphi}=\bsfvec{e}_{\varphi}\times{\nabla}A_{\varphi}'.
\label{eq:ernstpot1}
\end{equation}
Hereafter in this section the differential operators are to be understood as
three dimensional operators expressed in cylindrical coordinates in the
Euclidean space, $\bsfvec{e}_{\varphi}$ is a unit azimuthal vector, ${\rm
i}=\sqrt{-1}$. The E-M equations then turn into a pair of complex
equations~\cite{ernst:1968b}
\begin{equation}
\eqalign{
\left(\mathrm{Re}\,{\cal E}+\left|\Phi\right|^2\right)\Delta{\cal E} &=
\left({\nabla}{\cal E} +
2\Phi^{\ast}{\nabla}\Phi\right)\cdot{\nabla}{\cal E},\cr
\left(\mathrm{Re}\,{\cal E}+\left|\Phi\right|^2\right)\Delta\Phi &=
\left({\nabla}{\cal E} +
2\Phi^{\ast}{\nabla}\Phi\right)\cdot{\nabla}\Phi.\cr}
\label{eq:emeqernst}
\end{equation}

It is not difficult to verify, that the Ernst
equations~(\ref{eq:emeqernst}) are fulfilled for the ``charged''
$\gamma$-metrics~(\ref{eq:dsgamel}) and~(\ref{eq:dsgammg}) when one sets
\begin{equation}
\eqalign{\Phi &= A_{t} =
\frac{qf_{1}\left(r,z\right)}{1-q^2f_{1}\left(r,z\right)},\cr
{\cal E} &=
\frac{f_{1}\left(r,z\right)}{\left[1-q^2f_{1}\left(r,z\right)\right]^2} -
\left|\Phi\right|^2 = \frac{1}{q}\,\Phi.}
\label{eq:ernstel}
\end{equation}
for the $\gamma$-metric with an electric field~(\ref{eq:dsgamel}) and
\begin{equation}
\eqalign{\Phi &= {\rm i}A_{\varphi}',\qquad
A_{\varphi}' = 2qz + q\gamma\left(R_{1} - R_{2}\right),\cr
{\cal E} &=
f_{1}\left(r,z\right)\left[1+\frac{q^2r^2}{f_{1}\left(r,z\right)}\right]^2 -
\left|\Phi\right|^2.}
\label{eq:ernstmg}
\end{equation}
for the $\gamma$-metric with a magnetic field~(\ref{eq:dsgammg}). While in
the former case $\Phi$ is straightforwardly determined by the vector
potential~(\ref{eq:vpgamel}), in the latter case one has to solve a pair of
coupled partial differential equations to get $A_{\varphi}'$;
substituting~(\ref{eq:vpgammg}) into~(\ref{eq:ernstpot1}) these equations
read as
\[
\frac{\partial A_{\varphi}'}{\partial r}= qr\,\frac{\partial }{\partial
  z}\left[\ln f_1\left(r,z\right)\right],\qquad
\frac{\partial A_{\varphi}'}{\partial z}= 2q - qr\,\frac{\partial }{\partial
  r}\left[\ln f_1\left(r,z\right)\right],
\]

Formulating our problem through Ernst potentials ${\cal E}$ and $\Phi$ we
can easily found all dual solutions, as the duality rotation is represented
by the transformation $\Phi\longrightarrow\Phi\exp({\rm i}\alpha)$, where
$\alpha$ is a real constant~\cite{ernst:1968b}. Thus, according
to~(\ref{eq:ernstmg}) our ``magnetic'' solution ~(\ref{eq:dsgammg}) has a
dual electric counterpart the vector potential of which has a non-zero
time-like component $A_t=A_{\varphi}'=2qz + q\gamma\left(R_{1} -
R_{2}\right)$. Using~(\ref{eq:ernstel}), a magnetic solution dual to the
metric~(\ref{eq:dsgamel}) then requires
$A_{\varphi}' = A_t =
qf_{1}\left(r,z\right)/\left[1-q^2f_{1}\left(r,z\right)\right]$.
Solving a set of equations~(\ref{eq:ernstpot1}) for $A_{\varphi}$ we obtain
a vector potential with the only non-zero azimuthal component
$A_{\varphi}=q\gamma\left(R_{1} -R_{2}\right)$. In this sense, our
distinguishing of ``electric'' and ``magnetic'' solutions of E-M equations
is certainly rather conventional: it always depends on the choice of the vector
potential.

\section{Limiting cases}
\label{sec:limcases}
\subsection{Vacuum solutions}
\label{sec:limcasesvac}
In the preceding section we have found two classes of E-M fields. It is well
known that some vacuum Weyl metrics (e.g. the Curzon metric) represent limiting
cases of the Darmois-Voorhees-Zipoy solution~(\ref{eq:fcegam}). Consequently,
we can explore limiting cases of the spacetimes~(\ref{eq:dsgamel})
and~(\ref{eq:dsgammg}). In their inspiring paper~\cite{herrera:1999}
Herrera et al. presented a fruitful point of view that enables us to
systematize the particular cases efficiently. The unifying idea is based on
the comparison of the corresponding generating Newtonian potentials of
considered Weyl metrics. 

\begin{figure}[h]
\centerline{\hbox{\includegraphics{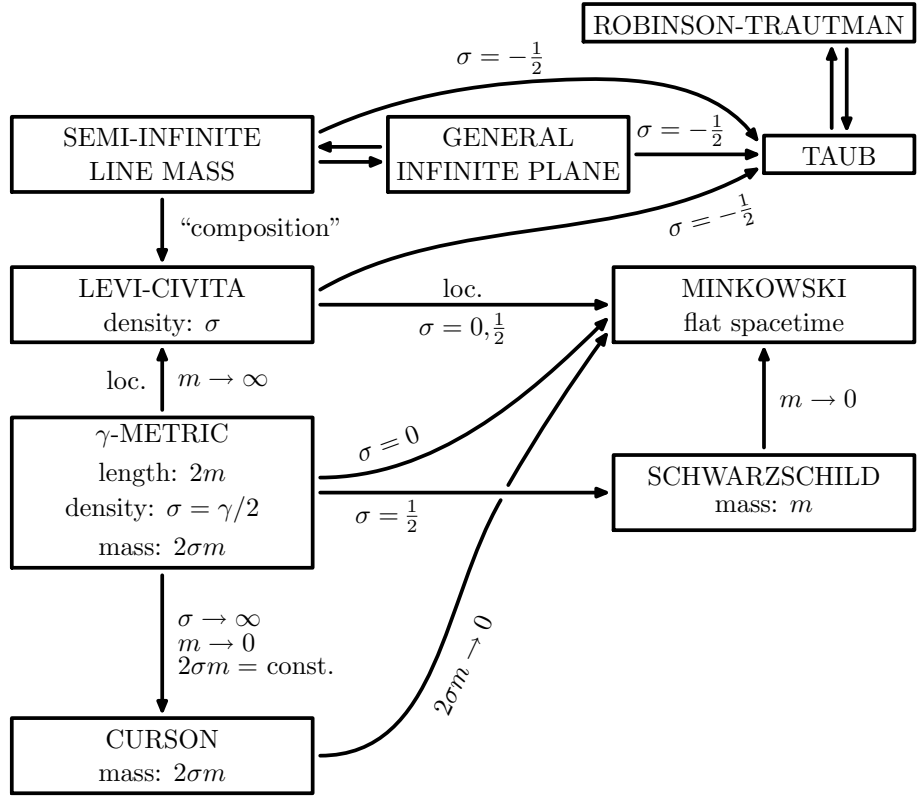}}}
\caption{Limiting diagram for the seed vacuum metrics from which E-M fields
can be generated through the H-M conjecture (the diagram is a generalization
of the similar one presented by Herrera et al.~\cite{herrera:1999})}.
\label{fig:fammet}
\end{figure}

The relations among the seed vacuum spacetimes are illustrated in
figure~\ref{fig:fammet}.
All these vacuum spacetimes represent axisymmetric solutions of Einstein's
equations or they can be brought into the Weyl form~(\ref{eq:dsweyl}) via a
suitable transformation. In the limiting diagram~\ref{fig:fammet} there are
two qualitatively different types of limits. The first one represents a
usual limit, when we set a particular value to some metric parameter (or
parameters); as an example we can take the Minkowski limit of the
$\gamma$-metric~(\ref{eq:dsgam}) for $\gamma=0$. The second type of limits
(denoted by the abbreviation ``loc.'' in the diagram) is achieved through
the Cartan scalars~\cite{herrera:1999}. Thus, such limit does not treat
global properties such as topological defects. To be concrete, setting
$\sigma=0$ in the line-element of the L-C metric~(\ref{eq:dslc}), we come to
the Minkowski metric in cylindrical coordinates, but the conical
singularity described by the parameter $C$ survives. Therefore, the
obtained limit globally differs from the Minkowski spacetime.

The choice of a coordinates in the limiting diagram follows Herrera et
al.~\cite{herrera:1999}, i.e. the Weyl-Lewis-Papapetrou cylindrical
coordinates of the L-C solution~(\ref{eq:dslc}) are introduced. The
coordinates in which the $\gamma$-metric takes the form~(\ref{eq:dsgam})
are just rescaled L-C coordinates with scaling ratios explicitly given
in~\cite{herrera:1999}.  Working with the L-C cylindrical coordinates, it
seems most convenient to express the parameters of other spacetimes through
the parameter $\sigma$ of the L-C solution~(\ref{eq:dslc}); the relations
given in the lower part of figure~\ref{fig:fammet} are also derived
in~\cite{herrera:1999}.

While relativistic limits sometimes require rather tedious calculations,
the relations among considered spacetimes become quite transparent when we
think about their generating Newtonian potentials. Thus, the relations
among the limits of the $\gamma$-metric provide a supporting argument for
the interpretation of the studied spacetimes according to the corresponding
Newtonian gravitational fields though e.g. in the case of L-C
solution~(\ref{eq:dslc}) such interpretation is acceptable only for a
limited range of the linear density parameter $\sigma$~\cite{bonnor:1991}.

In that sense, starting from the $\gamma$-metric~(\ref{eq:dsgam}), the
Newtonian image of which is a field of a finite rod laid along the
$z$-axis, and prolonging the rod at both ends to $-\infty$ and $+\infty$,
i.e. mathematically in the limit $m\to\infty$, we get a Newtonian field of
an infinite line-mass, the Newtonian analogy of the L-C
solution~(\ref{eq:dslc}). If, on the contrary, one contracts the length of the
rod to zero at the same time keeping its mass $M=2\sigma m=\gamma m$ finite
we obtain the Curzon metric, another Weyl metric~(\ref{eq:dsweyl}) with functions
$\mu$ and $\nu$ in the form~\cite{bonnor:1992}
\begin{equation}
\mu(r,z) = -\,\frac{M}{R},\quad 
\nu(r,z) = -\,\frac{M^2r^2}{2R^4},\quad
R = \sqrt{r^2+z^2}.
\label{eq:fcecurs}
\end{equation}
Its generating Newtonian potential $\phi\equiv\mu(r,z)$ describes the
gravitational field of a spherical particle.  The Curzon metric is again
endowed with a directional singularity and for a distant observer it looks
like a gravitational fields of a point particle with multipoles on
it~\cite{bonnor:1992} (let us remind that this vacuum solution can also
arise as a gravitational field of counter-rotating relativistic
discs~\cite{bicak:2000,bicak:1993}). Another limit of the $\gamma$-metric
for $\gamma=1$ is the well-known Schwarzschild spacetime, as it has been
already mentioned in section~\ref{sec:gammet}.  The $\gamma$-metric, the
Curzon metric as well as the Schwarzschild solution reduce to the Minkowski
spacetime when the mass of their source falls down to zero.

Let us concentrate on the upper part of the diagram which includes the seed
vacuum solutions studied thoroughly by Bonnor who has also found most
of the corresponding coordinate
transformations~\cite{bonnor:1992,bonnor:1991,bonnor:1990}.  An infinite
linear source, the Newtonian image of the vacuum L-C
solution~(\ref{eq:dslc}), can be considered as a ``composition'' of two
semi-infinite linear sources, each of which generates the Weyl metric
\begin{equation}
\eqalign{%
\displaystyle{{\rm d}s^{2}= -\,X^{2\sigma}{\rm d}t^{2} +
X^{-2\sigma}\left[\left(\frac{X}{2R}\right)^{4\sigma^{2}}\left({\rm d}r^{2}+
{\rm d}z^{2}\right)+r^{2}{\rm d}\varphi^{2}\right]},\\
X=R+\epsilon (z-z_{1}),\qquad R=\sqrt{r^{2}+(z-z_{1})^{2}}, 
}
\label{eq:dsseminflvac}
\end{equation}
The source of the generating Newtonian potential, a semi-infinite line-mass
with linear density $\sigma$ is again located along the $z$-axis either
from $z_1$ to $\infty$ ($\epsilon=-1$) or from $-\infty$ to $z_1$
($\epsilon=1$). As it has been pointed by
Bonnor~\cite{bonnor:1992,bonnor:1991}, the solution~(\ref{eq:dsseminflvac})
is isometric with a solution for a non-uniform infinite plane 
\begin{equation}
{\rm d}s^{2}=-\,Z^{4\sigma}{\rm d}t^{2} +\varrho^{2}Z^{2-4\sigma}{\rm
d}\varphi^{2} + Z^{8\sigma^{2}-4\sigma}\left(Z^{2} + \varrho^{2}\right)^{1
- 4\sigma^{2}} \left({\rm d}\varrho^{2}+{\rm d}Z^{2}\right);
\label{eq:dsinfplvac}
\end{equation}
the corresponding coordinate transformation~\cite{bonnor:1992,bonnor:1991}
maps~(\ref{eq:dsseminflvac}) into the half-space $Z>0$. In the
particular case $\sigma=-1/2$ all the solutions~(\ref{eq:dslc}),
(\ref{eq:dsseminflvac}), (\ref{eq:dsinfplvac}) are isometric with the Taub metric
\begin{equation}
{\rm d}s^{2}= 
\frac{1}{\sqrt{\xi}}\left(-{\rm d}\tau^{2}+{\rm d}\xi^{2}\right) + 
\xi\left({\rm d}x^{2}+{\rm d}y^{2}\right),
\label{eq:dstaub}
\end{equation}
the general plane symmetric solution~\cite{bonnor:1992,bonnor:1990}. 
Moreover, the metric~(\ref{eq:dstaub}) can be further transformed
into~\cite{bonnor:1990}
\begin{equation}
{\rm d}s^{2}= -\,\frac{2\alpha}{p}\,{\rm d}\eta^{2}-2\,{\rm d}p\,{\rm d}\eta +
p^{2}\left({\rm d}x^{2} + {\rm d}y^{2}\right),\qquad\alpha=\mbox{const.},
\label{eq:dsrobtr}
\end{equation}
a particular case of radiative vacuum solutions discovered by Robinson and
Trautman in~1962. The metric~(\ref{eq:dsrobtr}) is usually interpreted as a gravitational
field of a particle on a null line (properties of Robinson-Trautman's
solutions are summarized e.g. in~\cite{bicak:2000} with rather skeptic
conclusion about the cosmological and astrophysical relevance of these
solutions). This ambiguity illustrates the difficulty in physical
interpretation of the seed spacetimes listed in figure~\ref{fig:fammet} and
consequently, the problems with the interpretation of our generated E-M
fields.

The lower part of the figure~\ref{fig:fammet} has been described by Herrera
at al.~\cite{herrera:1999}. We do not intend to concentrate on the
exploration of these limits and coordinate transformations. From our point
of view all the vacuum spacetimes in the limiting diagram are interesting
from another reason: we have managed to apply the generalized H-M
conjecture to \emph{all} these seed metrics successfully and so we have
obtained several classes of E-M fields.  The existence of the limits and
coordinate transformations \emph{ex post} justifies the application of the
H-M conjecture to all these vacuum spacetimes according to the algorithmic
scheme formulated in~\cite{richterek:2000}. Some of the obtained limiting
E-M fields have been found before by other authors while the others are new
at least in Weyl cylindrical coordinates. The limiting diagram enables us
to systematize our earlier results because some known E-M fields may be
considered as special cases of the metrics~(\ref{eq:dsgamel})
or~(\ref{eq:dsgammg}).

The rest of this section is devoted to a brief survey of E-M fields
generated from the seed metrics in the limiting diagram. As the method
through which the E-M solutions were found was described
in~\cite{richterek:2000} and demonstrated in section~\ref{sec:gammetelmg},
we just list the line-elements ${\rm d}s^2$, vector potentials
$\bsfvec{A}$, Petrov types and -- in case of already known solutions --
relevant references. Validity of both Einstein and sourceless Maxwell's
equations can be easily checked by means of a suitable computer algebra
program.

\subsection{E-M fields of Curzon's type}\nopagebreak
\label{sec:em-curz-metr}
Evidently, one obtains these limits from equations~(\ref{eq:dsgamel}),
(\ref{eq:dsgammg}) when redefining the functions $f_1(r,z)$ and $f_2(r,z)$
according to relations~(\ref{eq:fcecurs}).

\smallskip\noindent{\em a) The Curzon solution with an electric field}\newline\nopagebreak
It represents a limit of the solution~(\ref{eq:dsgamel}) and its line
element has the form~(\ref{eq:dsweyl}) with
\[
f_1(r,z) = \exp\left(-\,\frac{2M}{R}\right),\
f_2(r,z) = \exp\left(-\,\frac{M^2r^2}{R^4}\right),\
f(r,z) = 1-q^2\exp\left(-\,\frac{2M}{R}\right).
\]
and the vector potential~(\ref{eq:vpgamel}) reduces into
\[
\bsfvec{A} = q\,\frac{1}{\exp\left(\,2M/R\right)-q^2}\,\bsfvec{dt}. 
\]

\smallskip\noindent{\em b) The Curzon solution with a magnetic field}\newline\nopagebreak
Similarly, this limit of the solution~(\ref{eq:dsgammg}) again represents
the Weyl metric~(\ref{eq:dsweyl}) with the choice
\[
\eqalign{%
f_1(r,z) = \exp\left(-\,\frac{2M}{R}\right),\quad
f_2(r,z) = \exp\left(-\,\frac{M^2r^2}{R^4}\right),\\
f(r,z) = 1+q^2r^2\exp\left(\frac{2M}{R}\right),\quad
\bsfvec{A} = q\,\frac{r^2}{\exp\left(-2M/R\right)+r^2q^2}\,\bsfvec{d}\bvphi. 
}
\]

\subsection{E-M fields of Levi-Civita's type}\nopagebreak
\label{sec:em-levciv}
Next two solutions has been found in~\cite{richterek:2000}. In the context
of this paper and especially of the limiting diagram in
figure~\ref{fig:fammet} they can be considered as limits of the
metrics~(\ref{eq:dsgamel}) (\ref{eq:dsgammg}) when one
extends $m\to\infty$ keeping $\sigma=\gamma/2$ constant.

\smallskip\noindent{\em a) The L-C solution with a radial electric field}\newline\nopagebreak
\label{sec:elw}
This E-M field represents a limit of~(\ref{eq:dsgamel}), its line element
and fourpotential read as
\begin{equation}
\eqalign{%
{\rm d}s^{2}&=-\frac{r^{4\sigma}}{f(r)^{2}}{\rm d}t^{2} +
f(r)^{2}r^{4\sigma (2\sigma - 1)}\left[{\rm d}r^{2} + {\rm
d}z^{2}\right] + f(r)^{2}C^{-2}r^{2-4\sigma}{\rm d}\varphi^{2},\\
f(r)&=1-q^2r^{4\sigma},\quad
\bsfvec{A} = -\frac{qr^{4\sigma}}{f(r)}\bsfvec{dt}.}
\label{eq:dselw}
\end{equation}
The metric (\ref{eq:dselw}) represents a special case of
the general class of cylindrically symmetric solution with radial electric
field given in \cite{KSMH},~\S 20.2, equation~20.9c. The line-element of
this general class has the form
\[
{\rm d}s^{2} = \varrho^{2m^2}G^{2}\,\left({\rm d}\varrho^2+{\rm
d}z^2\right) + \varrho^{2}G^{2}{\rm d}\varphi^{2} - G^{-2}{\rm d}t^{2},
\]
where $G=C_{1}\varrho^{m}+C_{2}\varrho^{-m}$ and $C_{1},C_{2},m$ are real
constants. The solution~(\ref{eq:dselw}) belongs generally to the Petrov type $I$, for
$\sigma=\pm1/2$  to the Petrov type $D$, for $\sigma=0$ the spacetime becomes
flat (Petrov type $0$).

\smallskip\noindent{\em b) The L-C solution with a longitudinal magnetic field}\newline\nopagebreak
\label{sec:mgwz}
Similarly, the limit of the metric~(\ref{eq:dsgammg}) takes the form
\begin{equation}
\eqalign{
{\rm d}s^{2}&=-f(r)^{2}r^{4\sigma}{\rm d}t^{2} + f(r)^{2}r^{4\sigma
(2\sigma - 1)}\left[ {\rm d}r^{2} + {\rm d}z^{2}\right] +
\frac{r^{2-4\sigma}}{f(r)^{2}C^{2}}\,{\rm d}\varphi^{2}.\\
f(r)&=1+\frac{q^{2}}{C^2}\,r^{2(1-2\sigma )},\qquad
\bsfvec{A}=\frac{qr^{2(1-2\sigma )}}{C^2f(r)}\,\bsfvec{d}\bvphi.}
\label{eq:dsmgz}
\end{equation}
Submitting the metric~(\ref{eq:dsmgz}) to the coordinate transformation
\begin{equation}
\varrho = r^{(2\sigma-1)^2}, \quad m=\frac{1}{2\sigma - 1},
\label{eq:maccaltr}
\end{equation}
one can show that the the solution~(\ref{eq:dsmgz}) belongs to the class of
the general static cylindrically symmetric solution with longitudinal
magnetic field~\cite{KSMH},~\S 20.2, equation~20.9b
\[
{\rm d}s^{2} = \varrho^{2m^2}G^{2}\,\left({\rm d}\varrho^2-{\rm
d}t^2\right) + G^{-2}{\rm d}\varphi^{2} + \varrho^{2}G^{2}{\rm d}z^{2},
\]
where $G=C_{1}\varrho^{m}+C_{2}\varrho^{-m}$ and $C_{1},C_{2},m$ are real
constants. The spacetime~(\ref{eq:dsmgz}) belongs to the Petrov type $I$,
for $\sigma=0,1$ to the type $D$ and for $\sigma=1/2$ it becomes
flat. In the case $\sigma=0$ the metric is equivalent to the Bonnor-Melvin
universe~(\ref{eq:dsbonmel}) (see also subsection~\ref{sec:em-mink}).

\subsection{E-M fields of Taub's type}
\label{sec:em-taub}
As it is apparent from the limiting diagram, these solutions are just the
spacetimes~(\ref{eq:dselw}) and~(\ref{eq:dsmgz}) in different coordinates
for a particular case $\sigma=-1/2$. To simplify analytic expressions a new
parameter $Q$ characterizing the strength of E-M field is introduced
below. It differs from the parameter $q$ used above (namely in
(\ref{eq:dselw}) and (\ref{eq:dsmgz})) only by some multiplicative
constant, the coordinate transformation can be found in~\cite{bonnor:1990}.

\smallskip\noindent{\em a) The Taub solution with an electric field.}\newline\nopagebreak
The L-C solution with radial electric field~(\ref{eq:dselw}) with
$\sigma=-1/2$ transforms into the one-parameter class of solutions
\begin{equation}
\eqalign{
{\rm d}s^{2}=&\displaystyle{-\frac{{\rm d}\tau^{2}}{\sqrt{\xi}
\left(1 - \displaystyle{\frac{Q^2}{\sqrt{\xi}}}\right)^{2}}  
+ \left(1-\displaystyle{\frac{Q^2}{\sqrt{\xi}}}\right)^{2}
\left[\frac{{\rm d}\xi^{2}}{\sqrt{\xi}}
+\xi\left( {\rm d}x^{2}+{\rm d}y^{2} \right)\right]}\\
&\bsfvec{A}=\displaystyle{-\,\frac{Q}{\sqrt{\xi}\left(1 - 
\displaystyle{\frac{\strut Q^{2}}{\sqrt{\xi}}}\right)}\,
\bsfvec{d}\btau, \quad Q = 2^{-2/3}\,q;}
}
\label{eq:taubrel}
\end{equation}
belonging to the Petrov type $D$.

\smallskip\noindent{\em b) The Taub solution with a magnetic field.}\newline\nopagebreak
Analogously, the L-C solution with longitudinal magnetic
field~(\ref{eq:dsmgz}) with $\sigma=-1/2$ transforms into the one parameter class
\begin{equation}
\eqalign{
{\rm d}s^{2}=&
\displaystyle{\left(1+Q^2\xi\right)^{2}\left[\frac{1}{\sqrt{\xi}}\left(-{\rm
d}\tau^{2}+{\rm d}\xi^{2}\right) + \xi\,{\rm d}y^{2}\right]+\frac{\xi\,{\rm
d}x^{2}}{(1+Q^{2}\xi)^2}}\\
&\bsfvec{A}=\displaystyle{\frac{Q\,\xi}{1+Q^{2}\xi}\,\bsfvec{dx},\quad Q = 2^{4/3}\,q;}
}
\label{eq:taubzmf}
\end{equation}
belonging to the Petrov type $I$.

\subsection{E-M fields of Robinson-Trautman's type}\nopagebreak
\label{sec:em-robtraut}
Next two one-parameter classes again do not represent new solution as they
can be obtained from the metrics of the Levi-Civita's
type~(\ref{eq:dselw}),~(\ref{eq:dsmgz}) setting $\sigma=-1/2$. For the sake
of simplicity a new rescaled constant parameter $Q$ for the strength of the
E-M field is again introduced as in the preceding subsection.

\smallskip\noindent{\em a) The Robinson-Trautman subclass with an electric
  field.}\newline\nopagebreak
Submitting the L-C solution with a radial electric field~(\ref{eq:dselw})
to the Bonnor's transformation~\cite{bonnor:1990}, one obtains the metric
\begin{equation}
\eqalign{
{\rm d}s^{2}=&\displaystyle{-\,\frac{2\alpha}{pf(p)^{2}}\,{\rm
d}\eta^{2}-\frac{2}{f(p)^{2}}\,{\rm d}p\,{\rm d}\eta +
p^{2}f(p)^{2}\,\left({\rm d}x^{2} + {\rm d}y^{2}\right)+}\\ 
&\displaystyle{+\,\frac{{\rm d}p^2}{f(p)^2}\left[Q^2\left(\frac{2}{\alpha}\right)^{2/3} +
\frac{3}{2}\frac{Q^4}{p}\left(\frac{2}{\alpha}\right)^{1/3} -
\frac{Q^6}{p^2}+\frac{Q^8}{8p^3}\left(2\alpha\right)^{1/3}\right]}\\
&f(p)=\displaystyle{1+Q^2\left(\frac{\alpha}{2}\right)^{1/3}\,\frac{1}{p},\quad
\bsfvec{A} = \displaystyle{-\frac{Q}{(4\alpha)^{1/3}f(p)}\,
\left(2\alpha\,\bsfvec{d}\blceta+p\,\bsfvec{dp}\right)},
\quad Q=q}
}
\label{eq:robtrelw}
\end{equation}
belonging to the Petrov type $D$.

\smallskip\noindent{\em b) The Robinson-Trautman subclass with a magnetic
  field.}\newline\nopagebreak 
Similarly, the L-C solution with a longitudinal magnetic
field~(\ref{eq:dsmgz}) is equivalent to the Petrov class $I$ metric
\begin{equation}
\eqalign{
{\rm d}s^{2}=&\displaystyle{-\,\frac{2\alpha}{p}\,f(p)^{2}\,{\rm
d}\eta^{2}-2f(p)^{2}\,{\rm d}p\,{\rm d}\eta + p^{2}\left[\frac{{\rm
d}x^{2}}{f(p)^{2}} + f(p)^{2}\,{\rm d}y^{2}\right]}\\
&f(p)=\displaystyle{1+Q^2p^2,\quad
\bsfvec{A}=\displaystyle{\frac{Qp^2}{f(p)}\,\bsfvec{dx}},
\quad Q=\left(\frac{2}{\alpha}\right)^{1/3}\frac{q}{C}}.}
\label{eq:robtrmgz}
\end{equation}

\subsection{General plane solution with an electromagnetic field}\nopagebreak
\label{sec:em-genpl}
The vacuum solution~(\ref{eq:dsinfplvac}) representing a non-uniform
infinite plane~\cite{bonnor:1992} has not been proved to be a limiting case
either of the $\gamma$-metric~(\ref{eq:dsgam}) nor the L-C
solution~(\ref{eq:dslc}). Nevertheless, the H-M conjecture can be
successfully applied to~(\ref{eq:dsinfplvac}) in the simplified way described
in~\cite{richterek:2000} and provides following E-M fields. 

\smallskip\noindent{\em a) An infinite plane with an electric field.}\newline\nopagebreak
Employing the timelike Killing vector $\partial_{t}$ of the seed
metric~(\ref{eq:dsinfplvac}), the application of the H-M conjecture leads to
the metric
\begin{equation}
\eqalign{
{\rm d}s^{2}=&\,f(Z)^{2}\left[\varrho^{2}Z^{2-4\sigma}{\rm d}\varphi^{2}
        +Z^{8\sigma^{2}-4\sigma}\left(Z^{2}+\varrho^{2}\right)^{1-4\sigma^{2}}\left({\rm
        d}\varrho^{2}+{\rm d}Z^{2}\right)\right]-\\
        &-\,\frac{Z^{4\sigma}}{f(Z)^{2}}\,{\rm d}t^{2},\quad
f(Z)=1-q^2Z^{4\sigma},\quad \bsfvec{A}=\frac{qZ^{4\sigma}}{f(Z)}\,\bsfvec{dt}.
}
\label{eq:dsinfplel}
\end{equation}
This solution belongs generally to the Petrov type $I$, for $\sigma=\pm1/2$
it reduces to the type $D$ and for $\sigma=0$ we get the Petrov type $0$.

\smallskip\noindent{\em b) An infinite plane with a magnetic field.}\newline\nopagebreak
Starting from the second Killing vector of the seed vacuum
solution~(\ref{eq:dsinfplvac}) $\partial_{\varphi}$, we come to the
spacetime
\begin{equation}
\eqalign{
{\rm d}s^{2}=&-\,\displaystyle{Z^{4\sigma}f(Z,\varrho)^{2}\,{\rm d}t^{2}
+\frac{\varrho^{2}Z^{2-4\sigma}}{f(Z,\varrho)^{2}}\,{\rm d}\varphi^{2}}+\\
&+\,f(Z,\varrho)^{2}Z^{8\sigma^{2}-4\sigma}\left(Z^{2}
+\varrho^{2}\right)^{1-4\sigma^{2}}
\left({\rm d}\varrho^{2}+{\rm d}Z^{2}\right),\\
&f(Z,\varrho)=\,1+q^2\varrho^{2}Z^{2-4\sigma},\quad
\bsfvec{A}=\frac{q\varrho^{2}Z^{2-4\sigma}}{f(Z,\varrho)}\,\bsfvec{d}\bvphi,
}
\label{eq:dsinfplmg}
\end{equation}
which generally belongs to the Petrov type $I$, for particular values
$\sigma=0,1/2,1$ to the Petrov type $D$.

\subsection{Semi-infinite line-mass with an electromagnetic field}\nopagebreak
\label{sec:em-seminflm}
The vacuum solution~(\ref{eq:dsseminflvac}) is isometric with the
metric~(\ref{eq:dsinfplvac})~\cite{bonnor:1992}. Thus, transforming the
metrics~(\ref{eq:dsinfplel}), (\ref{eq:dsinfplmg}) we straightforwardly
obtain the following E-M fields. 

\smallskip\noindent{\em a) Semi-infinite linear source with an electric
  field.}\newline\nopagebreak%
The class of E-M fields~(\ref{eq:dsinfplel}) transforms into
\begin{equation}
\eqalign{
{\rm d}s^{2}=&-\,\frac{X^{2\sigma}}{f(z,r)^2}\,{\rm d}t^{2} +
f(z,r)^2X^{-2\sigma}\left[\left(\frac{X}{2R}\right)^{4\sigma^{2}}
\left({\rm d}r^{2}+{\rm d}z^{2}\right)+r^{2}{\rm d}\varphi^{2}\right],\\
&f(z,r)=1-q^2X^{2\sigma},\quad 
\bsfvec{A}=\frac{qX^{\strut 2\sigma}}{f(z,r)}\bsfvec{dt}.
}
\label{eq:dssmilel}
\end{equation}

\smallskip\noindent{\em b) Semi-infinite linear source with a magnetic field}\newline\nopagebreak
Analogously, transforming the metric~(\ref{eq:dsinfplmg}), we come to the magneto-vacuum
E-M field
\begin{equation}
\eqalign{
{\rm d}s^{2}=&-\,f(z,r)^2\left[X^{2\sigma}{\rm d}t^{2} +
X^{-2\sigma}\left(\frac{X}{2R}\right)^{4\sigma^{2}}\left({\rm d}r^{2}+
{\rm d}z^{2}\right)\right] + 
\frac{r^{2}}{X^{2\sigma}f(z,r)^2}\,{\rm d}\varphi^{2},\\
&f(z,r)=1+q^2r^2X^{-2\sigma},\quad 
\bsfvec{A}=\frac{qr^{2}X^{\strut -2\sigma}}{f(z,r)}\bsfvec{d}\bvphi.
}
\label{eq:dssmilmg}
\end{equation}
The Petrov types are, of course, the same as in subsection~\ref{sec:em-genpl}.

\subsection{E-M fields of Schwarzschild's type}
\label{sec:em-schwarz}
We obtain these limits of both the solutions~(\ref{eq:dsgamel})
and~(\ref{eq:dsgammg}) straightforwardly setting $\gamma=1$
(i.e. $\sigma=1/2$) and introducing Erez-Rosen
coordinates~(\ref{eq:trerros}).

\smallskip\noindent{\em a) The Schwarzschild metric with a radial electric
  field.}\newline\nopagebreak 
Under above specified conditions the metric~(\ref{eq:dsgamel}) reduces to
\begin{equation}
\eqalign{
{\rm d}s^2=&-\,\frac{(1-2m/\varrho)}{\left(1-q^2+2q^2m/\varrho\right)^2}\,{\rm d}t^2 +\\
&+\left(1-q^2+2q^2m/\varrho\right)^2\left[\frac{{\rm d}\varrho^2}{1-2m/\varrho} +
\varrho^2\left({\rm d}\vartheta^2 + \sin^2\vartheta\,{\rm
d}\varphi^2\right)\right]
}
\label{eq:dsschwel}
\end{equation}
and the vector potential~(\ref{eq:vpgamel}) to
\[
\bsfvec{A} = q\frac{(1-2m/\varrho)}{1-q^2+2q^2m/\varrho}\,\bsfvec{dt}.
\]
Introducing a new radial coordinate
$R=\varrho\left(1-q^2+2q^2m/\varrho\right)$ and rescaling the time
coordinate $\tau=t/\left(1-q^2\right)$, we come to a familiar
Reissner-Nordstr\"om solution, which has been explored in connection with
the H-M conjecture in~\cite{horsky:1989}. The mass of the
Reissner-Nordst\"om source $M$ as well as its electric charge $Q$ depend
both on $m$ and $q$
\[
M=m\left(1+q^2\right),\quad Q=2mq;
\]
the vector potential then takes the form
\[
\bsfvec{A} = q\left(1-\frac{2m}{R}\right)\,\bsfvec{d}\btau.
\]
Evidently, in the vacuum case $q\to 0$ one gets the Schwarzschild solution as $M\to m$ and
$Q\to 0$ and $R\to\varrho$.

\smallskip\noindent{\em b) The Schwarzschild metric with a magnetic
  field.}\newline\nopagebreak
Analogously, the metric~(\ref{eq:dsgammg}) reduces to
\begin{equation}
\eqalign{
{\rm d}s^2=&\left(1+q^2\varrho^2\sin^2\vartheta\right)^2\left[
-\left(1-\frac{2m}{\varrho}\right)\,{\rm d}t^2 + \frac{{\rm d}r^2}{1-2m/\varrho} +
\varrho^2{\rm d}\vartheta^2\right] +\\
&+\frac{\varrho^2\sin^2\vartheta}{\left(1+q^2\varrho^2\sin^2\vartheta\right)^2}\,{\rm
d}\varphi^2
}
\label{eq:dsschwmg}
\end{equation}
and the vector potential~(\ref{eq:vpgammg}) to
\[
\bsfvec{A} = \frac{q\varrho^2\sin^2\vartheta}{1+q^2\varrho^2\sin^2\vartheta}\,\bsfvec{d}\bvphi.
\]

\subsection{Minkowski's limit}
\label{sec:em-mink}
The last two limits represent undoubtedly the simplest cases of all the E-M
fields listed in this section. They can be obtained either from the fields
of the Schwarzschild type~(\ref{eq:dsschwel}), (\ref{eq:dsschwmg}) when putting
$m=0$ and returning to the cylindrical coordinates $z=\varrho\cos\theta$,
$r=\varrho\sin\vartheta$, or from the E-M fields of the Levi-Civita
type~(\ref{eq:dselw}), (\ref{eq:dsmgz}) setting $\sigma=0$. In the
latter case the solutions might inherit a conical singularity of the
Levi-Civita solution determined by the constant $C$. Hereafter we assume $C=1$.

While the ``electric'' case is trivial as it is equivalent to the Minkowski
spacetime itself without any electromagnetic field, the solution with a
magnetic field has a well known form
\begin{equation}
{\rm d}s^{2}= \left(1+q^2r^{2}\right)^{2}
\left[ -{\rm d}t^{2} + {\rm d}r^{2} + {\rm d}z^{2}\right] +
\frac{r^{2}}{\left(1+q^2r^{2}\right)^2}\,{\rm d}\varphi^{2}.
\label{eq:dsbonmel}
\end{equation}
Evidently, this metric with $q=B_{0}/2$ represents so called Bonnor-Melvin
magnetic universe (see e.g.~\cite{KSMH}, \S 20.2, equation
20.10). As it was mentioned in subsection~\ref{sec:ernstpot}, the
spacetime~(\ref{eq:dsbonmel}) need not include a longitudinal magnetic
field only -- the background electromagnetic field might be also electric
(a particular example of this metric with a longitudinal electric field was found
in~\cite{richterek:2000}).

\section{Conclusion}
\label{sec:concl}
We have again proved explicitly that H-M conjecture outlined serves as a
useful, efficient tool for the generating of E-M fields. Let us briefly
summarize obtained results.

\begin{itemize}
\item[(i)] We present a successful application of the H-M conjecture in its
generalized formulation to the whole class of seed vacuum metrics. In each
case two Killing vectors of the corresponding seed vacuum solution were
used to generate new E-M fields. Thus, an algorithmic scheme for the
H-M conjecture formulated in~\cite{richterek:2000} can be evidently used in more
various situations, namely, in case of all vacuum solutions listed in
figure~\ref{fig:fammet}.
\item[(ii)] We have obtained several E-M fields generated from the seed vacuum
spacetimes of Weyl's type. Most of them represent a special case or a limit
of the general classes of E-M fields~(\ref{eq:dsgamel}) and
(\ref{eq:dsgammg}), some solutions can be derived through an appropriate
coordinate transformations. Components of basic tensors are explicitly
given and general Petrov classes are determined for all metrics.
\item[(iii)] It is typical for the solutions generated through the H-M conjecture
that their interpretation is determined by the physical features of their
seed vacuum metrics. Unfortunately, for most vacuum metrics discussed
above, several possible, often qualitatively different interpretations can
be proposed. However, Bonnor's interpretation~\cite{bonnor:1992,bonnor:1991}
seems to be most convenient for the discussion of the limits in
section~\ref{sec:limcases}. Therefore, we prefer to interpret the seed Weyl metrics
according to their generating Newtonian potentials. E-M fields obtained via
the H-M conjecture then might represent a gravitational field of linear
sources with some electric charge distribution or the linear sources in
a background electromagnetic field of the Bonnor-Melvin's type.
\end{itemize}

\subsubsection*{Acknowledgement}
The authors thank Prof. Malcolm A.~H.~MacCallum (Queen Mary \& Westfield
College, University of London) for drawing their attention to
the existence of the coordinate transformation~(\ref{eq:maccaltr}).
This work was supported by an internal grant of the Palack\'y University,
Olomouc, Czech Republic.

\appendix
\section*{Appendix: Circular geodesics in the equatorial plane}
\label{sec:circgeod}
The study of circular (azimuthal) geodesic can impose additional constraints
on the spacetime parameters. Hereafter we use Rindler-Perlick
method~\cite{rindler:1990} designed for axially symmetric stationary
metrics. The angular velocity $\omega={\rm d}\varphi/{\rm d}t$ along azimuthal
geodesics in the equatorial plane $z=0$ for any Weyl
solution~(\ref{eq:dsweyl}) was calculated by Herrera and
Pastora~\cite{herrera:2000} and in the Weyl-Lewis-Papapetrou {\em spherical}
coordinates it reads as
\begin{equation}
  \label{eq:omega}
  \omega = {\displaystyle\frac{\left|g_{tt}\right|
  \displaystyle\left(\frac{\partial\left|g_{tt}\right|}{\partial R}\right)^{1/2}}
  {\left(2R \left|g_{tt}\right| -
  R^2\,\displaystyle\frac{\partial\left|g_{tt}\right|}{\partial
  R}\right)^{1/2}}}.
\end{equation}
Applying~(\ref{eq:omega}) to our ``electric'' solution~(\ref{eq:dsgamel}) we obtain
\begin{equation}
\eqalign{\omega&=\frac{\sqrt{m\gamma}\,\left(R_{\rm e}^2-m^2\right)^{\gamma}}
{r\,\left[\left(R_{\rm e}+m\right)^{\gamma}-q^2\left( R_{\rm
        e}-m\right)^{\gamma}\right]^2}\,\times\\
&\times\,\sqrt{\frac{\left(R_{\rm e}+ m\right)^{\gamma}+q^2\left(R_{\rm
e}-m\right)^{\gamma}}{\left(R_{\rm e}-m\gamma\right)\left(R_{\rm e}+m\right)
^{\gamma}-q^2\left(R_{\rm e}+m\gamma\right)\left(R_{\rm
e}-m\right)^{\gamma}}}},
  \label{eq:omegael}
\end{equation}
analogously for our ``magnetic'' solution we get
\begin{equation}
\eqalign{\omega&=\frac{\left[q^2\,r^2\,\left(R_{\rm e}+m\right)^{\gamma}+\left(
 R_{\rm e}-m\right)^{\gamma}\right]^2\,}
{r\,\left(R_{\rm e}^2-m^2\right)^{\gamma}}\,\times\\
&\times\,\sqrt{\frac{q^2r^2\left(2
R_{\rm e}-m\gamma\right)\left(R_{\rm e}+m\right)^{\gamma}+\gamma\,m\,\left(R_{\rm e}
 -m\right)^{\gamma}}{\left(R_{\rm e}-\gamma\,m\right)
\left[\left(R_{\rm e}-m\right)^{\gamma}-q^2\,r^2\,\left(
 R_{\rm e}+m\right)^{\gamma}\right]}}},
  \label{eq:omegamg}
\end{equation}
where $R_{\rm e} = \left.R_1\right|_{z=0} = \left.R_2\right|_{z=0} =
\sqrt{r^2+m^2}$; in the equatorial plane both spherical and cylindrical radial
coordinates $R$, $r$ coincide. It is worth checking that in the vacuum
Schwarzschild limit ($q=0$ and $\gamma=1$) we come to the relation
$$
\omega^2={{m \left(\sqrt{r^{2}+m^{2}}-m\right)}\over{r^{2} \left(\sqrt{r^{
 2}+m^{2}}+m\right)^{2}}},
$$
which expressed in Erez-Rosen coordinates~(\ref{eq:trerros}) leads to the
``exact Kepler law'' $\omega^2=m\varrho^{-3}$ for the Schwarzschild
solution (it is derived e.g. in~\cite{rindler:1990},
$\varrho=\sqrt{(r^2+m^2)}+m$). The
expressions~(\ref{eq:omega})--(\ref{eq:omegamg}) would be much more
complicated for non-equatorial planes $z\neq 0$.

\begin{figure}[htbp]
  \centering
  \parbox[b]{.48\textwidth}{\centering%
  \includegraphics[width=0.48\textwidth]{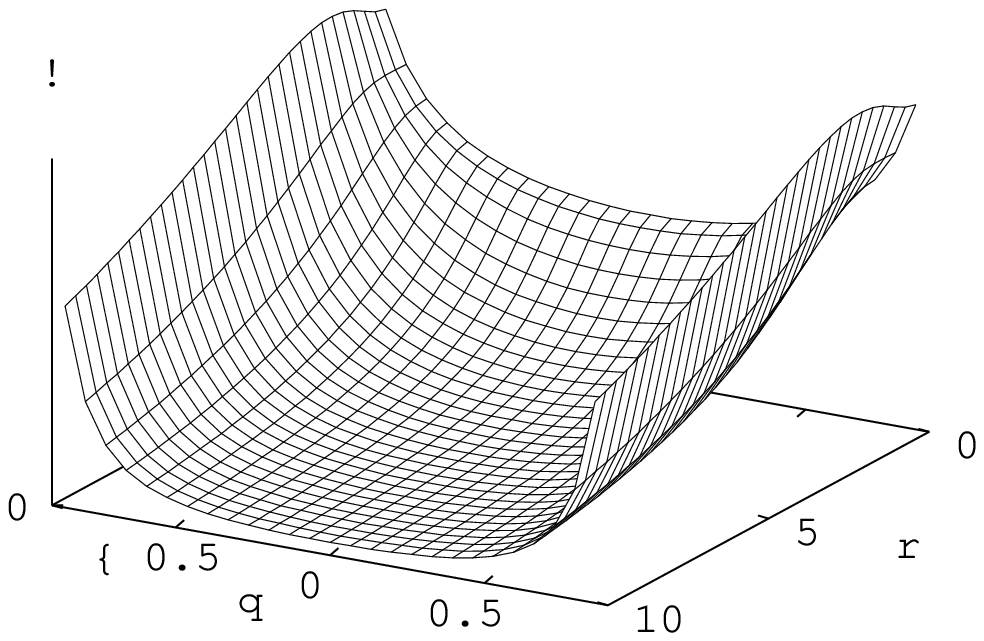}\par
  a)}\hfill
  \parbox[b]{.48\textwidth}{\centering%
  \includegraphics[width=0.48\textwidth]{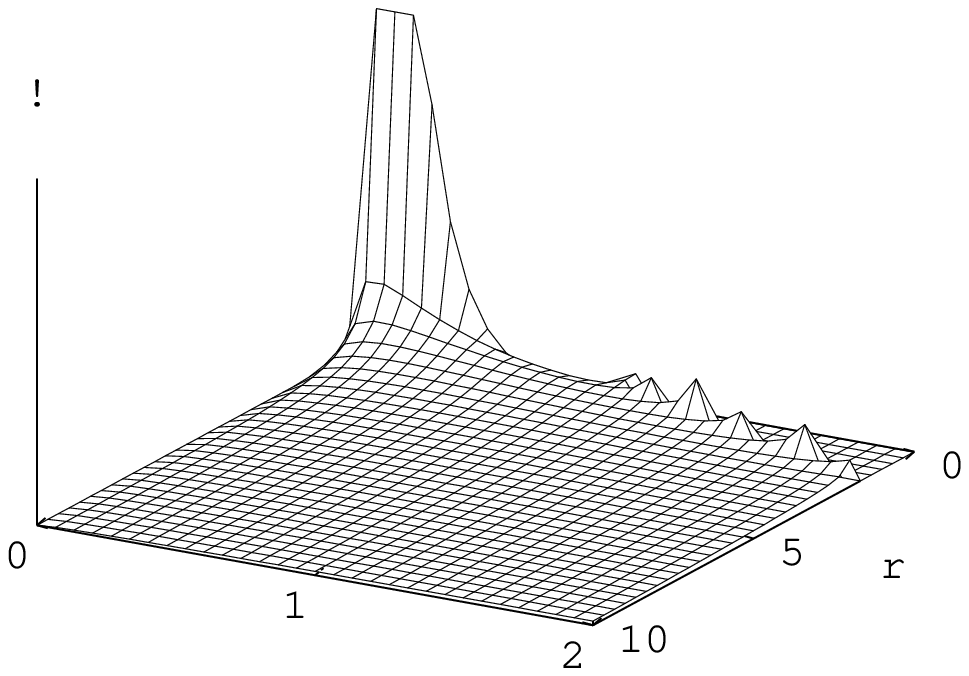}\par
  b)} 
  \caption{The angular velocity for the azimuthal geodesics in the
    equatorial plane of the solution~(\ref{eq:dsgamel}): a) $\gamma=1$; b) $q=0.5$.}
  \label{fig:omel}
\end{figure}

\begin{figure}[htbp]
  \centering
  \parbox[b]{.48\textwidth}{\centering%
  \includegraphics[width=0.48\textwidth]{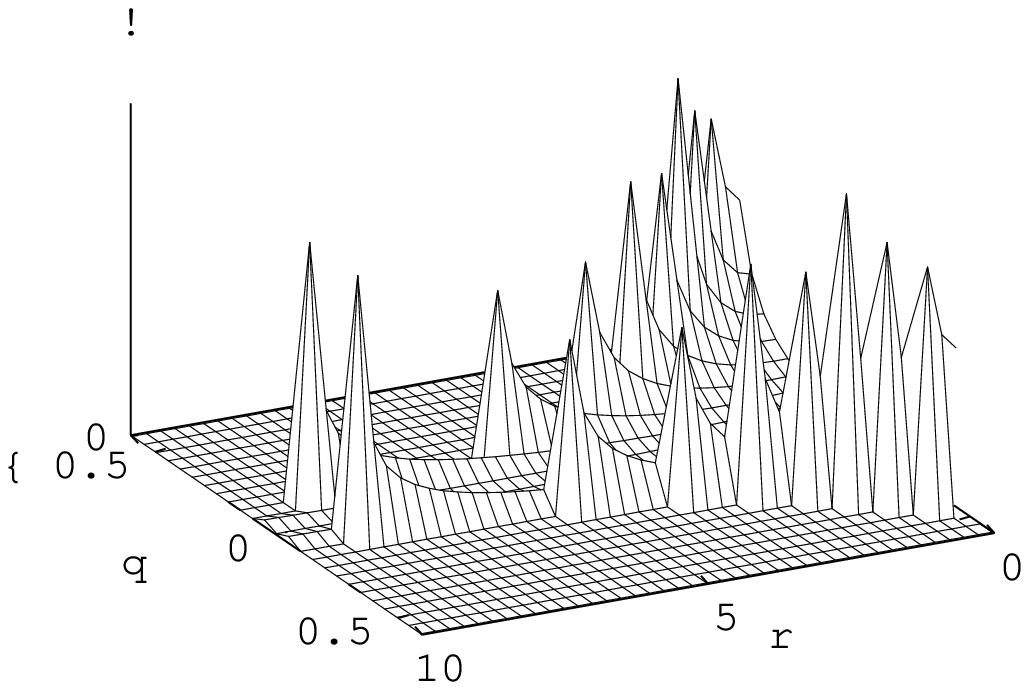}\par
  a)}\hfill
  \parbox[b]{.48\textwidth}{\centering%
  \includegraphics[width=0.48\textwidth]{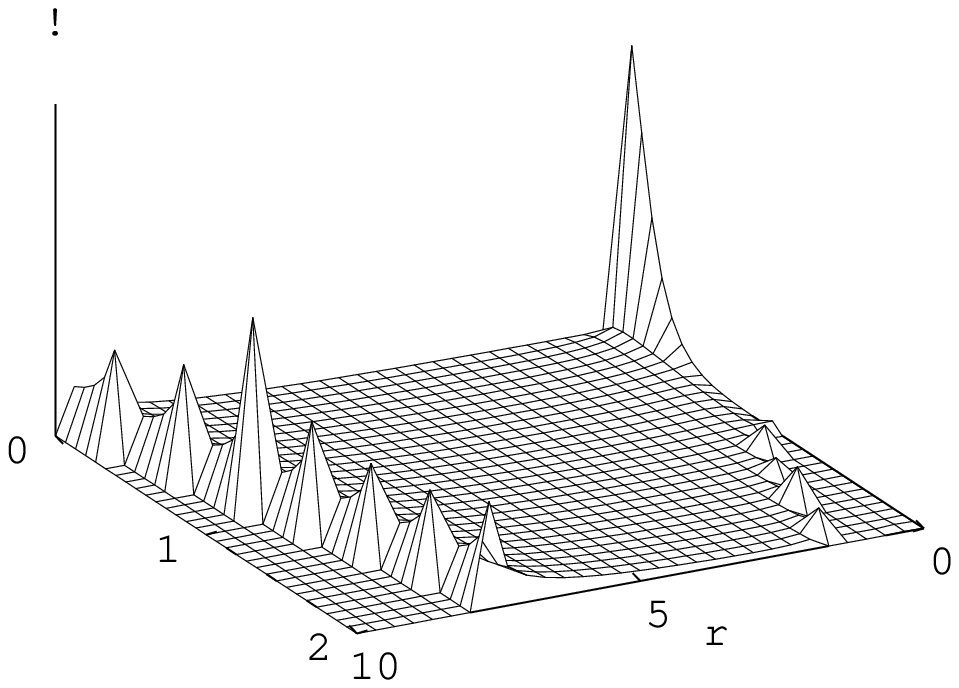}\par
  b)} 
  \caption{The angular velocity for the circular geodesics in the
    equatorial plane of the solution~(\ref{eq:dsgammg}): a) $\gamma=1$; b) $q=0.1$.}
  \label{fig:ommg}
\end{figure}

The dependence $\omega=\omega\left(r,q\right)$ for the~(\ref{eq:dsgamel}) with
  $\gamma=1$ (the ``charged'' Schwarzschild spacetime) is illustrated in
  figure~\ref{fig:omel}a. The circular geodesics are allowed only for
  $q\in\left(-1,1\right)$ and $\omega$ increases towards the symmetry axis
  $r=0$. According to the figure~\ref{fig:omel}b, this conclusion is valid
  for $\gamma\leq 1$, but for $\gamma>1$ the angular velocity $\omega$
  diverges at ``points'' $r\neq0$ and no azimuthal geodesics can be found
  in the region with sufficiently small $r$.
In case of the ``magnetic'' spacetime~(\ref{eq:dsgammg})
the existence of circular geodesics is limited not only at the side of
small $r$, but also at the side of large $r$, which is demonstrated in
figure~\ref{fig:ommg}.


\begin{thebibliography}{10}

\bibitem{KSMH}
D.~Kramer, H.~Stephani, E.~Herlt, and M.~A.~H. MacCallum:
{\em Exact Solutions of Einstein's Field Equations}.  Cambridge University
  Press,  Cambridge,
1980.

\bibitem{horsky:1989}
J.~Horsk\'y and N.~V. Mitskievitch:
Czech. J. Phys. {\rm B} \textbf{39} (1989)~957.

\bibitem{richterek:2000}
L.~Richterek, J.~Novotn\'y, and J.~Horsk\'y:
Czech. J. Phys. \textbf{50} (2000)~925
{\tt [gr-qc/0003004]}.

\bibitem{bonnor:1992}
W.~B. Bonnor:
Gen. Rel. Grav. \textbf{24} (1992)~551.

\bibitem{herrera:1999}
L.~Herrera, F.~M. Paiva, and N.~O. Santos:
J. Math. Phys. \textbf{40} (1999)~4064
{\tt [gr-qc/9810079]}.

\bibitem{bicak:2000}
J.~Bi\v{c}\'ak:
in {\em Einstein Field Equations and Their Physical Implications (Selected
  essays in honour of Juergen Ehlers)}, Lecture Notes in Physics {\bf 540}
  (Ed.~B.~G. Schmidt).  Springer-Verlag,  Berlin-Heidelberg-New York,
2000
{\tt [gr-qc/0004016]}.

\bibitem{bicak:1993}
J.~Bi\v{c}\'ak, D.~Lynden-Bell, and J.~Katz:
Phys. Rev. D \textbf{47} (1993)~4334.

\bibitem{herrera:1998}
L.~Herrera, F.~M. Paiva, and N.~O. Santos: (1998)
{\tt [gr-qc/9812023]}.

\bibitem{herrera:2000}
L.~Herrera and J.~L.~H. Pastora:
J. Math. Phys. \textbf{41} (2000)~7544
{\tt [gr-qc/0010003]}.

\bibitem{letelier:1998}
P.~S. Letelier and S.~R. Oliveira:
Class. Quant. Grav. \textbf{15} (1998)~421
{\tt [gr-qc/9710122]}.

\bibitem{wang:1997}
A.~Z. Wang, M.~F.~A. da~Silva, and N.~O. Santos:
Class. Quant. Grav. \textbf{14} (1997)~2417
{\tt [gr-qc/961052]}.

\bibitem{stephani:2000}
H.~Stephani:
Ann. Phys. (Leipzig) \textbf{9} (2000)~\SI 168.

\bibitem{ernst:1968a}
F.~J. Ernst:
Physical Review \textbf{167} (1968)~1175.

\bibitem{ernst:1968b}
F.~J. Ernst:
Physical Review \textbf{168} (1968)~1415.

\bibitem{bonnor:1991}
W.~B. Bonnor and M.~A.~P. Martins:
Class. Quant. Grav. \textbf{8} (1991)~727.

\bibitem{bonnor:1990}
W.~B. Bonnor:
Wissenschaft. Z. der Friedrich-Schiller Universit\"at Jena \textbf{39}
  (1990)~25.

\bibitem{rindler:1990}
W.~Rindler and V.~Perlick:
Gen. Rel. Grav. \textbf{22} (1990)~1067.

\end{thebibliography}

\end{document}